\documentstyle[epsfig]{aipproc}

\begin{document}

\title{Environmental Influences in SGRs and AXPs}

\author{David Marsden$^*$, Richard Lingenfelter$^{\dagger}$, and 
Richard Rothschild$^{\dagger}$, and James Higdon$^{\ddag}$}
\address{$^*$NASA/Goddard Space Flight Center\thanks{NAS/NRC research 
associate}\\
Greenbelt, Md 20771\\
$^{\dagger}$Center for Astrophysics and Space Sciences, University 
of California at San Diego \\ La Jolla, CA 92093 \\
$^{\ddag}$W. M. Keck Science Center, Claremont Colleges, Claremont,
CA 91711}

\maketitle

\begin{abstract}

Soft gamma--ray repeaters (SGRs) and anomalous x--ray pulsars 
(AXPs) are young ($<$100 kyr), radio-quiet, x-ray pulsars which 
have been rapidly spun-down to slow spin periods clustered at 
$5-12$ s. Nearly all of these unusual pulsars also appear to be 
associated with supernova shell remnants (SNRs) with typical ages 
$<20$ kyr. If the unusual properties of SGRs and AXPs were due to 
an innate feature, such as a superstrong magnetic field, then the 
pre-supernova environments of SGRs and AXPs should be typical of 
neutron star progenitors. This is {\it not} the case, however, as 
we demonstrate that the interstellar media which surrounded the 
SGR and AXP progenitors and their SNRs were unusually dense 
compared to the environments around most young radio pulsars 
and SNRs. Thus, if these SNR associations are real, the SGRs 
and AXPs can not be ``magnetars'', and we suggest instead that 
the environments surrounding SGRs and AXPs play a controlling 
role in their development.

\end{abstract}

\section{Introduction}

Soft gamma--ray repeaters (SGRs) are neutron stars whose multiple 
bursts of gamma--rays distinguish them from other gamma--ray burst 
sources\cite{hurley00}. SGRs are also unusual x--ray pulsars in 
that they have spin periods clustered in the interval $5-8$ s, and 
they all appear to be associated with supernova remnants (SNRs), 
which limits their average age to approximately $20$ kyr\cite{braun89}. 
The angular offsets of the SGRs from the apparent centers of their 
associated supernova remnant shells indicates that SGRs are endowed 
with space velocities $>500$ km s$^{-1}$, which are greater than the 
space velocities of most radio pulsars\cite{cordes98}. Anomalous x--ray 
pulsars (AXPs) are similar to SGRs in that they are radio quiet x--ray 
pulsars with spin periods clustered in the range $6-12$ s, and have 
similar\cite{mer99} persistent x--ray luminosities as the SGRs 
($\sim 10^{35}$ ergs s$^{-1}$). Most of the AXPs appear to be 
associated with supernova remnants, and therefore they are also 
thought to be young neutron stars like the SGRs. Here we present 
a new look at environmental evidence which shows that the SGRs and 
AXPs can not be due to a purely innate property, such as superstrong 
magnetic fields\cite{thompson95}. 

\section{The Environments of SGRs and AXPs}
\label{environment}

If the unusual properties of SGRs and AXPs were due solely to an 
intrinsic property of the neutron star, that developed independently 
of the external environment, then the characteristics of the 
interstellar medium which surrounded the AXP and SGR progenitors 
should be typical of that around the massive O and B stars which 
are progenitors of all neutron stars. Observations clearly show 
that the majority of neutron stars are formed in ``superbubbles'': 
evacuated regions of the ISM which surround the OB associations 
in which the massive progenitors of most neutron stars live. The 
supernovae from the massive O and B stars which form SGRs and AXPs 
are heavily clustered in space and time and form vast ($>100$ 
pc) HII regions/superbubbles\cite{maclow88} filled with 
a hot ($>10^{6}$ K) and tenuous ($n\sim 10^{-3}$ cm$^{-3}$) 
gas. The occurrence of most supernovae in the hot phase of 
the ISM is confirmed from observations of nearby 
galaxies\cite{vandyk96} and from studies of Galactic 
SNRs\cite{higdon80}. It is estimated that $90\pm 10\%$ of 
all core-collapse supernova should occur in this hot and 
tenuous environment\cite{higdon98}.
 
The environments of SGRs and AXPs are probed by the blastwaves of 
their associated supernova remnants, and from the size of the remnant 
shell as a function of the age we can constrain the external density. 
In Table $1$ we have listed the $12$ known SGRs and AXPs and their 
associated supernova remnant shells\cite{marsden00}. The identification 
of the associated remnants are based on both positional coincidences 
of the remnant and the SGR/AXP, and on similar distances of the 
SGR/AXP and its associated remnant. We include the new tentative
\cite{cline99} SGR candidate 1801--23, which appears to be associated 
with the SNR W28. The thin SGR error box passes roughly through the 
center of the SNR and through the compact, nonthermal x--ray 
source\cite{andrews83} within the remnant. No associated remnants 
can be found for AXPs 0720--3125 and 0142$+$615, which is not surprising 
given the close distance ($\sim 0.1$) of 0720--3125\cite{haberl97}, 
and the molecular clouds associated with 0142$+$615\cite{israel94}. 
A more detailed discussion and reference list for the sources in Table 
$1$ will be published elsewhere\cite{marsden00}.  

Most of the SGR/AXP positions are significantly displaced from the 
apparent centers of their associated SNRs, as can be seen in Table 
$1$ from the ratio of the neutron star angular displacement 
$\theta_{\ast}$ divided by the angular radius $\theta_{SNR}$ of 
the remnant shell. These displacements clearly indicate that the 
SGR/AXPs have large transverse velocities. There is considerable 
uncertainty in the actual velocities, however, because the estimated 
remnant ages are probably uncertain by a factor of two in most cases, 
which introduces a corresponding uncertainty in the transverse 
velocities. In addition, the actual space velocities of the 
SGR/AXPs are larger by an unknown factor dependent on the viewing 
angle. Nonetheless, the data suggest that the typical SGR/AXPs 
are of the order of $1000$ km s$^{-1}$. Such velocities, while 
much larger than the typical neutron star velocities, are not 
unprecedented, as $\sim 10\%$ of radio pulsars may have space 
velocities of $1000$ km s$^{-1}$ or greater\cite{cordes98}. We 
conclude, therefore, that the SGRs and AXPs are a {\it high 
velocity} subset of young neutron stars. 

In Figure $1$ we have plotted the SNR shell radii as a function of 
the estimated age of each remnant. Overplotted in solid lines are simple 
approximations of the evolutionary tracks\cite{shull89} of supernova 
remnant expansion in the wide range of the external ISM densities, 
and we see that these SNRs are all in the denser ($>0.1$ cm$^{-3}$) 
phases of the ISM which slow their expanding shells to $<2000$ km 
s$^{-1}$ in $<10$ kyr. Also overplotted are the tracks of neutron 
stars born at the origin of the supernova explosion with varying 
velocities, showing the times required for fast (e.g. $>500$ km 
s$^{-1}$) neutron stars to catch up with the slowing supernova ejecta 
and swept-up matter.

\begin{figure}
%\vspace{-0.5truein}
\centerline{\epsfig{file=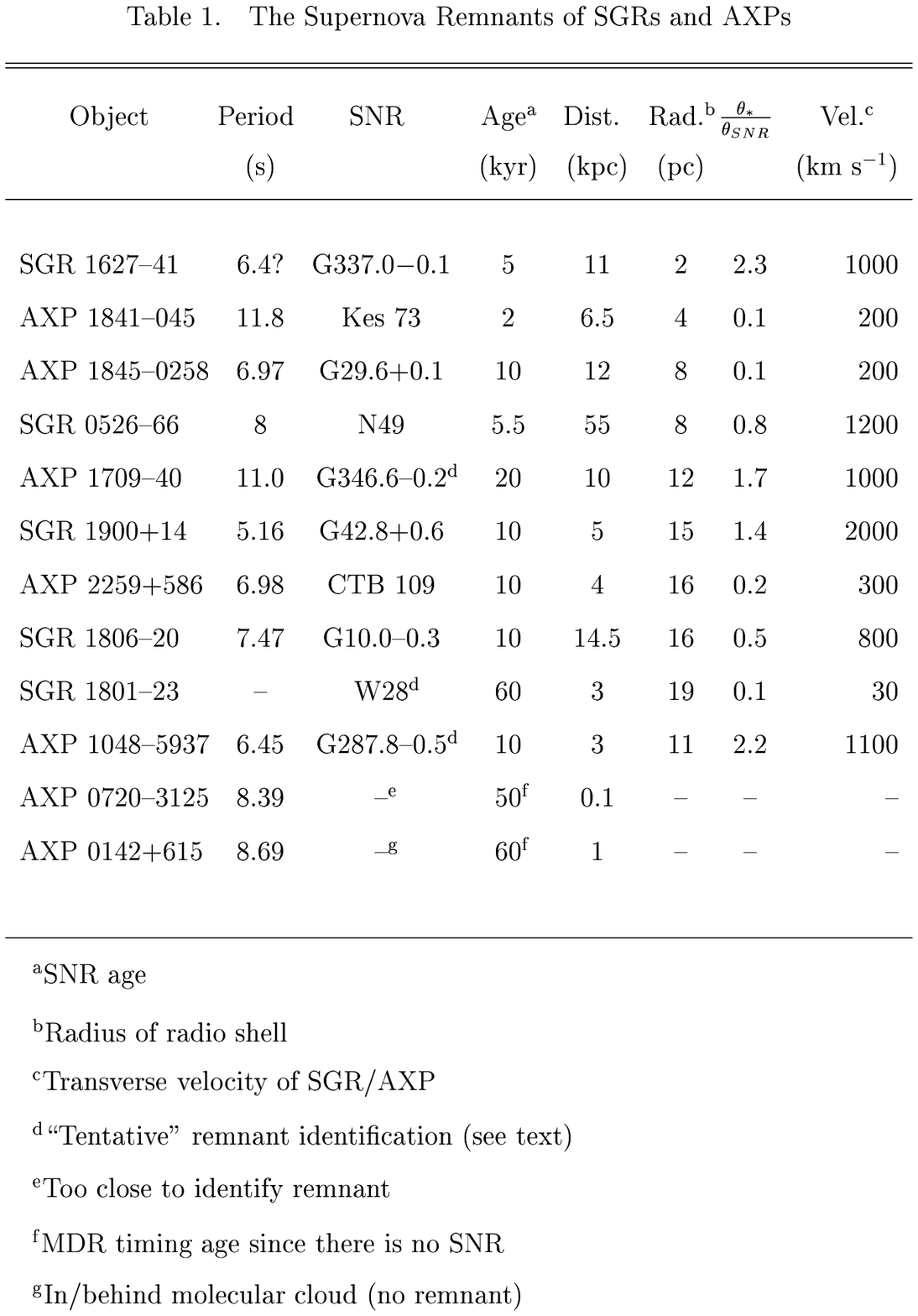,height=5.0in,width=4.0in}}
\vspace{-1.truein}
\end{figure}

\setcounter{figure}{1}
\begin{figure}
\vspace{-1.0truein}
\centerline{\epsfig{file=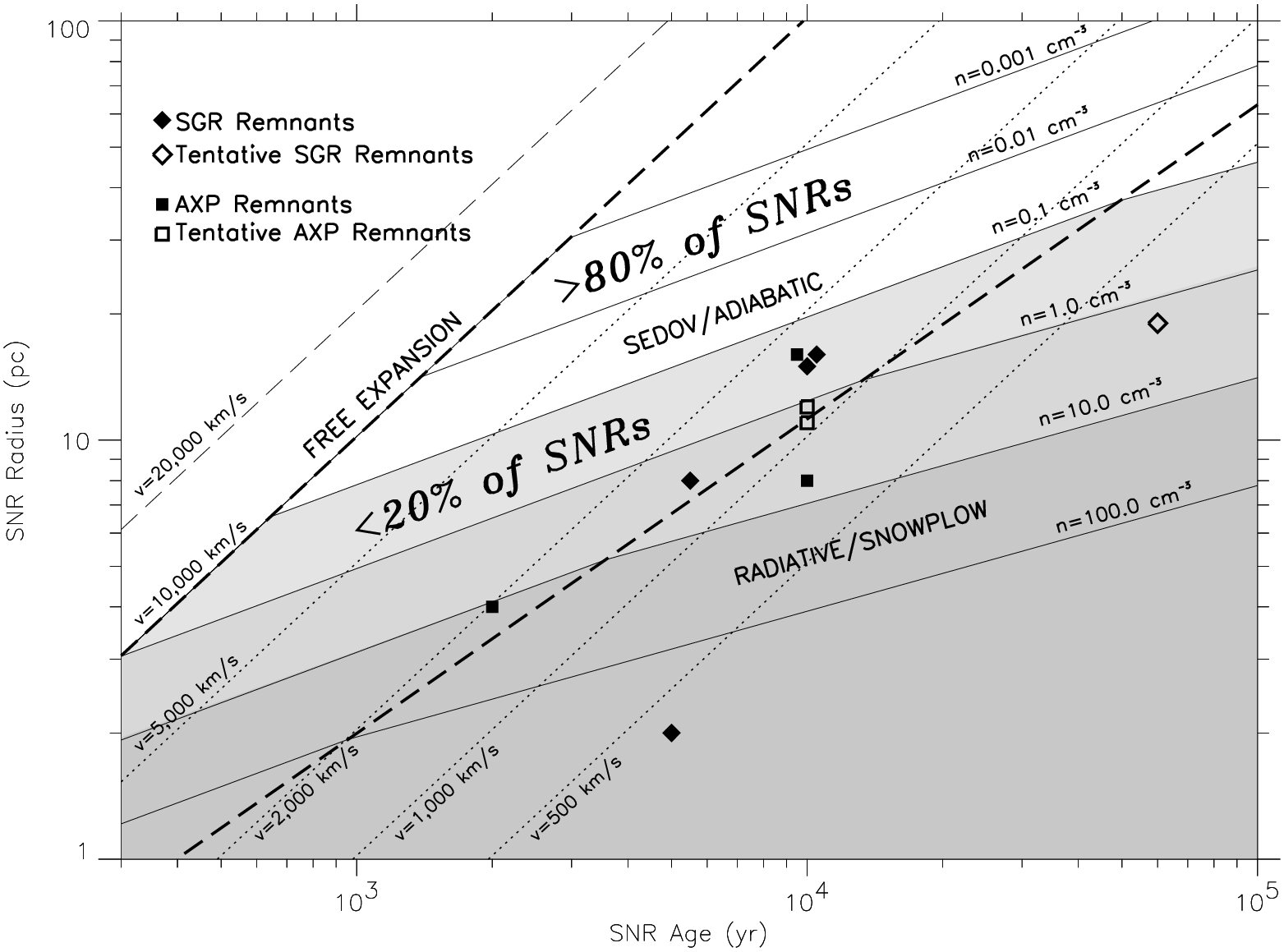,height=4in,width=3.5in}}
%\vspace{-0.1truein}
\caption{~The radius of the SGR and AXP supernova remnant shells 
as a function of their age. The solid lines denote SNR expansion 
trajectories and the dotted lines denote the tracks of neutron stars 
born at the origin of the supernova explosion with varying space 
velocities. We see that essentially all of these sources were formed 
in the denser phase of the interstellar medium (ISM), which clearly 
indicates that the environment, and not a purely intrinsic property 
such as a superstrong magnetic field, is the controlling factor in 
the development of the SGRs and AXPs}
\end{figure}

\section{Discussion}

From the discussion in $\S$ \ref{environment}, we saw that neutron 
stars should preferentially reside in the diffuse ($n< 0.01$ 
cm$^{-3}$) gas which constitutes the hot phase of the interstellar 
medium. As seen from Figure $1$, however, the SGRs and AXPs tend form 
in denser regions of the ISM. Given the entire sample of AXPs and 
SGRs, the probability that this is merely due to chance depends on 
the the ability to detect supernova remnants in the different phases 
of the interstellar medium. For the SGRs, the detection sensitivity 
is independent of the interstellar medium, because they are detected 
via their bright gamma--ray/x--ray bursts. Therefore, using only the 
SGRs yields a chance probability of less than $(0.2)^{5}\sim 10^{-4}$, 
if one accepts the tentative W28/SGR 1801--23 association, and $\sim 
10^{-3}$ if one excludes SGR 1801--23 from the SGR sample. The AXPs 
are also preferentially in the dense phase, which further lowers the 
chance probability for the class as a whole. The evidence then suggests 
that the environments surrounding SGRs and AXPs are significantly 
different than otherwise normal neutron stars in a way which is 
{\it inconsistent} with the hypothesis that the properties of these 
sources are the result of an innate characteristic such as a 
superstrong magnetic field. 

These observational facts imply instead that the environment is 
crucial in the development of SGRs and AXPs. One plausible scenario 
is that the rapid spin-down of the SGR/AXPs may result from 
their interaction with co-moving ejecta and swept-up ISM material
\cite{corbet95}\cite{vanparadijs95}. Calculations\cite{marsden00} 
indicate that such an {\it interaction} scenario, involving the 
formation of accretion disks by fast ($>500$ km s$^{-1}$) neutron 
stars from co-moving ejecta of supernova remnants slowed to $<2000$ 
km s$^{-1}$ by the denser ($>0.1$ cm$^{-3}$) phases of the ISM, could 
spin-down SGRs and AXPs to their present-day spin periods in $\sim 
10$ kyr -- consistent with the estimated ages of these sources -- 
without requiring the existence of a population of neutron stars 
with ultrastrong magnetic fields. In addition, such a scenario can 
explain the clustering of spin periods, present-day spin-down 
rates, and the number of SGRs and AXPs in our galaxy\cite{marsden00}.

\end{document}